# Functional Programming in Pattern-Match-Oriented Programming Style

Satoshi Egi[a,b] and Yuichi Nishiwaki[b]

a    Rakuten Institute of Technology, Rakuten, Inc., Japan
b    The University of Tokyo, Japan

**Abstract**    Throughout the history of functional programming, recursion has emerged as a natural method for describing loops in programs. However, there does often exist a substantial cognitive distance between the recursive definition and the simplest explanation of an algorithm even for the basic list processing functions such as `map`, `concat`, or `unique`; when we explain these functions, we seldom use recursion explicitly as we do in functional programming. For example, `map` is often explained as follows: the map function takes a function and a list and returns a list of the results of applying the function to all the elements of the list.

This paper advocates a new programming paradigm called *pattern-match-oriented* programming for filling this gap. An essential ingredient of our method is utilizing pattern matching for non-free data types. Pattern matching for non-free data types features non-linear pattern matching with backtracking and extensibility of pattern-matching algorithms. Several non-standard pattern constructs, such as not-patterns, loop patterns, and sequential patterns, are derived from this pattern-matching facility. Based on that result, this paper introduces many programming techniques that replace explicit recursions with an intuitive pattern by confining recursions inside patterns. We classify these techniques as pattern-match-oriented programming design patterns.

These programming techniques allow us to redefine not only the most basic functions for list processing such as `map`, `concat`, or `unique` more elegantly than the traditional functional programming style, but also more practical mathematical algorithms and software such as a SAT solver, computer algebra system, and database query language that we had not been able to implement concisely.

## The Art, Science, and Engineering of Programming

**Functional Programming in Pattern-Match-Oriented Programming Style**

## 1 Introduction

How do you answer the question, "What is the map function?" We believe most people answer as follows:

1. "The map function takes a function and a list and returns a list of the results of applying the function to all the elements of the list."

Few people answer as follows:

2. "The map function takes a function and a list and returns an empty list if the argument list is empty. Otherwise, it returns a list whose head element is the result of applying the function to the head element of the argument list, and the tail part is the result of applying the map function recursively to the tail part of the argument list."

Obviously, there is a significant gap between these two explanations. The former explanation is simpler and more straightforward than the latter. However, the current functional definition of map is based on the latter.

```
map _ [] = []
map f (x : xs) = (f x) : (map f xs)
```

Interestingly, this basic definition of map has been almost unchanged for 60 years since McCarthy first presented the definition of maplist in [16]. The only difference is a way for describing conditional branches: McCarthy uses predicates, whereas Haskell uses pattern matching.

```
maplist[x ;f] = [null[x] -> NIL; T -> cons[f[car[x]]; maplist[cdr[x]; f]]]
```

Recursion used in the above definitions is a mathematically simple but powerful framework for representing computations and has been a very basic construct of functional programming for describing loops in programs. Recursion is heavily used for definitions of many basic functions such as filter, concat, and unique and most of them are also simple enough.

However, as mentioned earlier, there *does* exist a substantial cognitive distance between the recursive definition and the simplest explanation of that definition. To illustrate this gap, we define mapWithBothSides, a variation of map. We often meet a situation to define a variant of basic functions specific to a target algorithm. Defining these utility functions is one of the cumbersome tasks in programming. Therefore, considering a comfortable method for defining these utility functions is important.

mapWithBothSides takes a function of three arguments and a list, and returns a list of applying the function for all three-tuples consisting of an initial prefix, the next element, and the remaining suffix. mapWithBothSides is used for generating lists by rewriting the element of an input list. This function is useful for handling logical formulae, for example. We define mapWithBothSides with a helper function as follows.

```
mapWithBothSides f xs = mapWithBothSides' f [] xs
 where
  mapWithBothSides' f xs [] = []
  mapWithBothSides' f xs (y : ys) = (f xs y ys) : (mapWithBothSides' f (xs ++ [y]) ys)
```

The explanation of `map` given at the beginning and the above explanation of `mapWithBothSides` are similar, and their definitions should be similar to each other. However, their definitions are very different from each other. A hint for filling the gap is hidden behind these differences.

The cause of these differences is the lack of a pattern like `hs ++ ts` that divides a target list into an initial prefix and the remaining suffix. For example, the list [1,2] has multiple decompositions for the pattern `hs ++ ts`: [] ++ [1,2], [1] ++ [2], and [1,2] ++ []. We call this pattern a *join pattern*. Unlike traditional pattern matching for algebraic data types, this pattern has multiple decompositions. Pattern matching that can handle multiple results is necessary for handling join patterns.

In Egison [23] whose distinguishing feature is non-linear pattern matching with backtracking [9], we can define `map` and `mapWithBothSides` concisely and in a very similar way. In fact, the creator of Egison, an author of this paper, got an idea of the language when he implemented `mapWithBothSides` for implementing an automated theorem conjecturer. By the way, in this paper, we present Egison in Haskell-like syntax to omit a detail explanation of the syntax.

```
map f xs = matchAll xs as list something with _ ++ $x : _ -> f x
mapWithBothSides f xs = matchAll xs as list something with $hs ++ $x : $ts -> f hs x ts
```

In the above program, the join pattern is effectively used. **matchAll** is a key built-in syntactic construct of Egison for handling multiple pattern-matching results. **matchAll** collects all the pattern-matching results and returns a collection where the body expression has been evaluated for each result. **matchAll** takes one additional argument *matcher* that is `list something` in the above cases. A matcher is an Egison specific object that knows how to decompose the target following the given pattern. The matcher is specified between **as** and **with**, which are reserved words. `list` is a user-defined function that takes a matcher for the elements and returns a matcher for lists. `list` defines a method for interpreting the cons (:) and join (++) pattern. `something` is the only built-in matcher in Egison. `something` can be used for pattern-matching arbitrary objects but can handle only pattern variables and wildcards. As a result, `list something` is evaluated as a matcher for pattern-matching a list of arbitrary objects. `_` that appears in a pattern is a wildcard. Pattern variables are prepended with $.

These definitions of the variations of `map` are close to the explanation of `map` given at the beginning. We achieved this by hiding the recursions in the definition of `list`, which defines the pattern-matching algorithm for the join patterns. We call this programming style that replaces explicit recursions with intuitive patterns, *pattern-match-oriented programming style*.[1]

The purpose of this paper is to advocate pattern-match-oriented programming as a new programming paradigm. Pattern-match-oriented programming is a paradigm that makes descriptions of algorithms concise by replacing loops for traversal and enumeration that can be done by simple backtracking. These loops are mixed in programs with loops that are essential for lowering the time complexity of algorithms and make programs complicated. Pattern-match-oriented programming separates these two kinds of loops and allows programmers to concentrate on describing the essential part of algorithms. This distinction of two kinds of loops is illustrated in a SAT solver example in section 4.1.

---

[1] Our approach that hides explicit recursions in higher-level abstraction is similar to that of recursion schemes [17]. What makes our approach unique is that we hide recursions in patterns, not in functions.

The above examples show just a part of the expressiveness of pattern-match-oriented programming. For example, non-linear patterns enable to describe more popular list processing functions such as `unique` and `intersect` in pattern-match-oriented style. Furthermore, pattern matching for general non-free data types (e.g., multisets, sets, and mathematical expressions) diversifies the applications of pattern-match-oriented programming significantly. This paper introduces full features of the Egison pattern-match-oriented programming language and presents all the techniques we discovered so far for replacing explicit recursions with an intuitive pattern.

The contributions of the paper are summarized as follows:

1. Advocate a new programming paradigm that replaces recursions with intuitive patterns, called pattern-match-oriented programming;
2. Show usefulness of a syntactic construct and several non-standard pattern constructs such as `matchAllDFS` (section 2.4), *not-patterns* (section 2.5), *loop patterns* (section 2.6), and *sequential patterns* (section 2.7) with many working examples;
3. Classify programming techniques utilizing the advanced pattern-matching facility as programming design patterns (section 3), and demonstrate their usefulness in practical situations (section 4).

The remainder of this paper is organized as follows. Section 2 introduces Egison and various pattern constructs for non-free data types. These pattern constructs increase the number of situations in which we can replace verbose recursions with more intuitive patterns. Section 3 catalogs pattern-match-oriented programming techniques utilizing the features introduced in section 2. Section 4 explores the effect of pattern-match-oriented programming in more practical situations. Section 5 reviews the related work. Finally, section 6 concludes the paper.

## 2 Quick Tour of the Egison Pattern-Match-Oriented Language

This section quickly introduces the pattern-matching facility of Egison. The features explained in the former part of this section (section 2.1, section 2.2, and section 2.3) are already discussed in the original paper of Egison [9]. The features explained in the latter part are newly introduced in this paper except for loop patterns that are intensively discussed in the author's previous paper [6].

### 2.1 Value Patterns and Predicate Patterns for Representing Non-linear Patterns

`matchAll` gets even more powerful when combined with non-linear patterns. For example, the following non-linear pattern matches when the target collection contains a pair of identical elements.

```
matchAll [1,2,3,2,4,3] as list integer with _ ++ $x : _ ++ #x : _ -> x -- [2,3]
```

Value patterns play an important role in representing non-linear patterns. A value pattern matches the target if the target is equal to the content of the value pattern. A value pattern is prepended with # and the expression after # is evaluated referring to the value bound to the pattern variables that appear on the left side of the patterns. As a result, for example, $x : #x : _ is valid, but #x : $x : _ is invalid.

Let us show pattern matching for *twin primes* as a sample of non-linear patterns. Twin primes are pairs of prime numbers whose forms are $(p, p+2)$. primes is an infinite list of prime numbers. This matchAll extracts all twin primes from this infinite list of prime numbers in order.

```
twinPrimes = matchAll primes as list integer with _ ++ $p : #(p + 2) : _ -> (p, p + 2)

take 10 twinPrimes -- [(3,5),(5,7),(11,13),(17,19),(29,31),(41,43),(59,61),(71,73),(101,103),(107,109)]
```

There are cases that we might want to use more general predicates in patterns than equality. Predicate patterns are provided for such a purpose. A predicate pattern matches the target if the predicate returns true for the target. A predicate pattern is prepended with ?, and a predicate of one argument follows after ?.

```
twinPrimes = matchAll primes as list integer with _ ++ $p : ?(\q -> q == p + 2) : _ -> (p, p + 2)
```

## 2.2 Non-linear Pattern Matching with Backtracking

The pattern-matching algorithm inside Egison includes the backtracking mechanism for efficient non-linear pattern matching.[2]

```
matchAll [1..n] as list integer with _ ++ $x : _ ++ #x : _ -> x
-- returns [] in O(n^2) time
matchAll [1..n] as list integer with _ ++ $x : _ ++ #x : _ ++ #x : _ -> x
-- returns [] in O(n^2) time
```

The above expressions match a collection that consists of integers from 1 to *n* as a list of integers for enumerating identical pairs and triples, respectively. This target collection contains neither identical pairs nor triples. Therefore both expressions return an empty collection.

When evaluating the second expression, Egison interpreter does not try pattern matching for the second #x because pattern matching for the first #x always fails. Therefore, the time complexities of the above expressions are identical. The pattern-matching algorithm inside Egison is discussed in [9] in detail.

---

[2] The Author's paper [8] on Scheme macros that implement our pattern-matching system discusses the performance of a program in pattern-match-oriented style. A program in pattern-match-oriented style is currently two times slower when compared with the same program in functional programming style.

**Functional Programming in Pattern-Match-Oriented Programming Style**

### 2.3 Ad-hoc Polymorphism of Patterns by Matchers

Another merit of matchers, in addition to the extensibility of pattern-matching algorithms, is the ad-hoc polymorphism of patterns. The ad-hoc polymorphism of patterns is important for non-free data types because some data are pattern-matched as various non-free data types at the different parts of a program. For example, a collection is pattern-matched as a list, a multiset, and a set. Polymorphic patterns reduce the number of names for pattern constructors.

In the following sample, a list [1,2,3] is pattern-matched using different matchers with the same cons pattern. In the case of multisets, the cons pattern decomposes a collection into an element and the rest elements ignoring the order of the elements. In the case of sets, the rest elements are the same as the original collection because we ignore the redundant elements. If we regard a set as a collection that contains infinitely many copies of each element, this specification of the cons pattern for sets is natural.

```
matchAll [1,2,3] as list integer with $x : $xs -> (x,xs) -- [(1,[2,3])]
matchAll [1,2,3] as multiset integer with $x : $xs -> (x,xs) -- [(1,[2,3]),(2,[1,3]),(3,[1,2])]
matchAll [1,2,3] as set integer with $x : $xs -> (x,xs) -- [(1,[1,2,3]),(2,[1,2,3]),(3,[1,2,3])]
```

Polymorphic patterns are useful especially for value patterns. As well as other patterns, the behavior of value patterns is dependent on matchers. For example, an equality [1,2,3] == [2,1,3] between collections is false if we regard them as lists but true if we regard them as multisets. Still, thanks to ad-hoc polymorphism of patterns, we can use the same syntax for both types. This dramatically improves the readability of the program and makes programming with non-free data types easy.

```
matchAll [1,2,3] as list integer with #[2,1,3] -> "Matched" -- []
matchAll [1,2,3] as multiset integer with #[2,1,3] -> "Matched" -- ["Matched"]
```

### 2.4 matchAllDFS for Controlling the Order of Pattern-Matching Process

The **matchAll** expression is designed to enumerate all countably infinite pattern-matching results. For this purpose, users sometimes need to care about the order of pattern-matching results.

Let us start by showing a representative sample. The **matchAll** expression below enumerates all pairs of natural numbers. We extract the first 8 elements with the take function. **matchAll** traverses the reduction tree of pattern matching in breadth-first search to traverse all the nodes (Sect. 5.2 of [9]). As a result, the order of the pattern-matching results is as follows.

```
take 6 (matchAll [1..] as set something with $x : $y : _ -> (x,y)) -- [(1,1),(1,2),(2,1),(1,3),(2,2),(3,1)]
```

The above order is preferable for traversing an infinitely large reduction tree. However, sometimes, this order is not preferable (see section 3.1.2 and section 3.4.1). **matchAllDFS** that traverses a reduction tree in depth-first order is provided for this reason.

```
take 6 (matchAllDFS [1..] as set something with $x : $y : _ -> (x,y)) -- [(1,1),(1,2),(1,3),(1,4),(1,5),(1,6)]
```

## 2.5 Logical Pattern Constructs: And-Patterns, Or-Patterns, and Not-Patterns

The situations where and-patterns and or-patterns are useful are similar to those of the existing languages, whereas not-patterns become useful when they are combined with non-linear pattern matching with backtracking.

We start by showing pattern matching for *prime triples* as an example of and-patterns and or-patterns. A prime triple is a triple of primes whose form is $(p, p+2, p+6)$ or $(p, p+4, p+6)$. The and-pattern is used as an as-pattern. The or-pattern is used to match both of $p+2$ and $p+4$.

```
primeTriples = matchAll primes as list integer with
        _ ++ $p : (and (or #(p + 2) #(p + 4)) $m) : #(p + 6) : _ -> (p, m, p + 6)

take 8 primeTriples -- [(5,7,11),(7,11,13),(11,13,17),(13,17,19),(17,19,23),(37,41,43),(41,43,47),(67,71,73)]
```

A not-pattern matches a target if the pattern does not match the target, as its name implies. A not-pattern is prepended with !, and a pattern follows after !. The following **matchAll** enumerates sequential pairs of prime numbers that are *not* twin primes.

```
take 10 (matchAll primes as list integer with _ ++ $p : (and !#(p + 2) $q) : _ -> (p, q))
-- [(2,3),(7,11),(13,17),(19,23),(23,29),(31,37),(37,41),(43,47),(47,53),(53,59)]
```

## 2.6 Loop Patterns for Representing Repetition

A loop pattern is a pattern construct for representing a pattern that repeats multiple times. It is an extension of Kleene star operator of regular expressions for general non-free data types [6].

Let us start by considering pattern matching for enumerating all combinations of two elements from a target collection. It can be written using **matchAll** as follows.

```
comb2 xs = matchAll xs as list something with _ ++ $x_1 : _ ++ $x_2 : _ -> [x_1, x_2]

comb2 [1,2,3,4] -- [[1,2],[1,3],[2,3],[1,4],[2,4],[3,4]]
```

Egison allows users to append indices to a pattern variable as $x_1 and $x_2 in the above sample. They are called *indexed variables* and represent $x_1$ and $x_2$ in mathematical expressions. The expression after _ must be evaluated to an integer and is called an *index*. We can append as many indices as we want like x_i_j_k. When a value is bound to an indexed pattern variable $x_i, the system initiates an abstract map consisting of key-value pairs if x is not bound to a map, and bind it to x. If x is already bound to a map, a new key-value pair is added to this map.

Now, we generalize `comb2`. The loop patterns can be used for that purpose.

```
comb n xs = matchAll xs as list something with
        loop $i (1,n)
          (_ ++ $x_i : ...)
          _ -> map (\i -> x_i) [1..n]

comb 2 [1,2,3,4] -- [[1,2],[1,3],[2,3],[1,4],[2,4],[3,4]]
comb 3 [1,2,3,4] -- [[1,2,3],[1,2,4],[1,3,4],[2,3,4]]
```

The loop pattern takes an *index variable, index range, repeat pattern*, and *final pattern* as arguments. An index variable is a variable to hold the current repeat count. An index range specifies the range where the index variable moves. An index range is a tuple of an *initial number* and *final number*. A repeat pattern is a pattern repeated when the index variable is in the index range. A final pattern is a pattern expanded when the index variable gets out of the index range.

Inside loop patterns, we can use the *ellipsis pattern* (...). The repeat pattern or the final pattern is expanded at the location of the ellipsis pattern. The repeat pattern is expanded replacing the ellipsis pattern incrementing the value of the index variable.

The repeat counts of the loop patterns in the above samples are constants. However, we can also write a loop pattern whose repeat count varies depending on the target by specifying a pattern instead of an integer as the final number. When the final number is a pattern, the ellipsis pattern is replaced with both the repeat pattern and the final pattern, and the repeat count when the ellipsis pattern is replaced with the final pattern is pattern-matched with that pattern. The following loop pattern enumerates all initial prefixes of the target collection.

```
matchAll [1,2,3,4] as list something with loop $i (1, $n) ($x_i : ...) _ -> map (\i -> x_i) [1..n]
-- [[],[1],[1,2],[1,2,3],[1,2,3,4]]
```

Loop patterns are heavily used especially for trees and graphs. We work on pattern matching for trees in section 3.4.1. More formal specification of syntax and semantics of loop patterns is shown in the author's previous paper [6].

### 2.7 Sequential Patterns for Controlling the Order of Pattern-Matching Process

The pattern-matching system of Egison processes patterns from left to right in order. However, there are cases where we want to change this order, for example, to refer to the value bound to the right side of a pattern. Sequential patterns are provided for such a purpose.

Sequential patterns allow users to control the order of the pattern-matching process. A sequential pattern is represented as a list of patterns. Pattern matching is executed for each pattern in order. In the following sample, the target list is pattern-matched from the third, first, and second element in order.

```
matchAll [2,3,1,4,5] as list integer with
 [ @ :    @    : $x : _,
  (#(x + 1), @),
  #(x + 2)] -> "Matched" -- ["Matched"]
```

@ that appears in a sequential pattern is called *later pattern variable*. The target data bound to later pattern variables are pattern-matched in the next sequence. When multiple later pattern variables appear, they are pattern-matched as a tuple in the next sequence. It allows us to apply not-patterns for different parts of a pattern at the same time as we will see in section 3.3.

Some readers might wonder that a sequential pattern can be transformed into a nested **matchAll** expression. There are at least two reasons why it is impossible. First, a nested **matchAll** expression breaks breadth-first search strategy: the inner **matchAll** for the second result of the outer **matchAll** is executed only after the inner **matchAll** for the first result of the outer **matchAll** is finished. Second, a later pattern variable retains the information of not only a target but also a matcher. There are cases that the matcher of **matchAll** is a parameter passed as an argument of a function, and a pattern is polymorphic. Therefore, it is impossible to determine the matchers of inner **matchAll** expressions syntactically.

**2.8 Matcher Compositions**

Matchers are composable. We can define matchers for such as tuples of multisets and multisets of multisets. Using this feature, we can define matchers for various data types.

First, we can define a matcher for tuples by a tuple of matchers. A tuple pattern is used for pattern matching using such a matcher. For example, we can define the intersect function using a matcher for tuples of two multisets. We work on pattern matching for tuples of collections more in section 3.3.

```
intersect xs ys = matchAll (xs,ys) as (multiset eq, multiset eq) with ($x : _, #x : _) -> x
```

eq is a user-defined matcher for data types for which equality is defined. When the eq matcher is used, equality is checked for a value pattern.[3]

By passing a tuple matcher to a function that takes and returns a matcher, we can define a matcher for various non-free data types. For example, we can define a matcher for a graph as a set of edges. In the following code, we assume a node id is represented by an integer.

```
graph = multiset (integer, integer)
```

A matcher for adjacency graphs also can be defined. An adjacency graph is defined as a multiset of tuples of an integer and a multiset of integers.

```
adjacencyGraph = multiset (integer, multiset integer)
```

Some readers might wonder about matchers for algebraic data types. Egison provides a special syntactic construct for defining a matcher for an algebraic data type. For example, a matcher for binary trees can be defined using algebraicDataMatcher.

```
algebraicDataMatcher binaryTree a = BLeaf a | BNode a (binaryTree a) (binaryTree a)
```

Matchers for algebraic data types and matchers for non-free data types also can be combined. For example, we can define a matcher for trees whose nodes have an arbitrary number of children whose order is ignorable. We show pattern matching for these trees in section 3.4.1.

```
algebraicDataMatcher tree a = Leaf a | Node a (multiset (tree a))
```

---

[3] A definition of the eq matcher is explained in Sect. 6.3 of [9].

**Functional Programming in Pattern-Match-Oriented Programming Style**

## 3 Pattern-Match-Oriented Programming Design Patterns

This section introduces basic pattern-match-oriented programming techniques that replace explicit recursions with intuitive patterns. In the first part of this section, we rewrite many list processing functions such as map, filter, elem, delete, any, every, unique, concat, and difference, for which we expect most functional programmers imagine the same definitions. In the latter part of this section, we move our focus to descriptions of more mathematical algorithms that are not well supported in the current functional programming languages. We proceed with this section by listing patterns that frequently appear and show situations in which they are useful. The following table shows this list.

| Name | Description | Explained and Used in |
| --- | --- | --- |
| Join-cons pattern for list | Enumerate combinations of elements. | 3.1 |
| Cons pattern for multiset | Enumerate permutations of elements. | 3.2, 3.3, 3.4, 4.1, 4.2, 4.3 |
| Tuple pattern for collections | Compare multiple collections. | 3.3, 3.4, 4.1, 4.3 |
| Loop pattern | Describe repetitions inside patterns. | 3.4 |

### 3.1 Join-Cons Patterns for Lists — List Basic Functions

Join patterns whose second argument is a cons pattern, such as _ ++ $x : _, are frequently used for lists. We call these patterns *join-cons patterns*. Many basic list processing functions can be redefined by simply using this pattern.

#### 3.1.1 Single Join-Cons Patterns — The map Function and Its Family

_ ++ $x : _ matches each element of the target collection when the list matcher is used. As a result, the **matchAll** expression below matches each element of xs, and returns the results of applying f to each of them. As discussed in Introduction, this map definition is very close to our natural explanation of map.

```
map f xs = matchAll xs as list something with _ ++ $x : _ -> f x
```

By modifying the above **matchAll** expression, we can define several functions. For example, we can define filter by inserting a predicate pattern.

```
filter pred xs = matchAll xs as list something with _ ++ (and ?pred $x) : _ -> x
```

We can define elem by using a value pattern. elem is a predicate that determines whether the first argument element appears in the second argument list or not. **match** is provided also in Egison. **match** is just an alias of head (**matchAll** ...) because Egison evaluates **matchAll** lazily.[4]

```
elem x xs = match xs as list eq with
       _ ++ #x : _ -> True
       _          -> False
```

---

[4] **matchAll** also can handle multiple match clauses. **matchAll** $t$ **as** $m$ **with** $c_1$ $c_2$ ... is equivalent to **matchAll** $t$ **as** $m$ **with** $c_1$ ++ **matchAll** $t$ **as** $m$ **with** $c_2$ ++ ....

We can define delete that removes the first appearance of x from xs by modifying elem.

```
delete x xs = match xs as list eq with
        $hs ++ #x : $ts -> hs ++ ts
        _               -> xs
```

The predicate any and every [22] also can be concisely defined with predicate patterns using **match**. any is a predicate that determines whether any element of the second argument list satisfies the first argument predicate. every is a predicate that determines whether all elements of the second argument list satisfy the first argument predicate.

```
any pred xs = match xs as list something with
        _ ++ ?pred : _ -> True
        _              -> False

every pred xs = match xs as list something with
        _ ++ !?pred : _ -> False
        _               -> True
```

### 3.1.2 Nested Join-Cons Patterns — The unique and concat Function

By combining multiple join-cons patterns, we can describe more expressive patterns. One example is the unique function. The unique function is defined in the pattern-match-oriented style as follows.

```
unique xs = matchAll xs as list eq with _ ++ $x : !(_ ++ #x : _) -> x
```

A not-pattern is used to describe that there is *no* more x after an occurrence of x. Therefore, this pattern extracts only the last appearance of each element.

```
unique [1,2,3,2,4] -- [1,3,2,4]
```

We can define unique whose results consist of the first appearance of each element by rewriting the above pattern using a predicate pattern with the elem predicate. To match only the first appearance of an element, we rewrite a pattern that ensures that the same element does not appear before that element. We cannot write such a pattern with a simple combination of the cons and join patterns because they match a target list from left to right.

```
unique xs = matchAll xs as list eq with $hs ++ (and !?(\x -> elem x hs) $x) : _ -> x
```

```
unique [1,2,3,2,4] -- [1,2,3,4]
```

Another more elegant solution is using a sequential pattern. We can describe the same pattern by using the sequential pattern for the first argument of join.

```
unique xs = matchAll xs as list eq with
        [@ ++ $x : _,
         !(_ ++ #x : _)] -> x
```

Another example of a nested join-cons pattern is concat. We can define concat in the pattern-match-oriented style by combining a nested join-cons pattern and matcher composition. Note that **matchAllDFS** is necessary for ordering the output list properly.

**Functional Programming in Pattern-Match-Oriented Programming Style**

```
concat xss = matchAllDFS xss as list (list something) with _ ++ (_ ++ $x : _) : _ -> x
```

If we used **matchAll** instead of **matchAllDFS** for concat, it enumerates the elements of the input list of lists alternately.

```
matchAll [[1..], (map negate [1..])] as list (list something) with _ ++ (_ ++ $x : _) : _ -> x
-- [1,2,-1,3,-2,4,-3,5,-4,6]
```

### 3.2 Cons Patterns for Multisets

Cons patterns for a multiset are useful when we want to treat a collection ignoring the order of elements. We often meet such a situation especially when describing mathematical algorithms.

We start from a simple example. The lookup function for association lists can be defined using a single cons pattern for multiset. A single cons pattern for a multiset can be replaced by a join-cons pattern for a list.

```
lookup k ls = match ls as multiset (eq, something) with (#k, $x) : _ -> x
```

The usage of cons patterns for multisets differs from that of join-cons patterns when they are nested. Cons patterns for multisets can be used to enumerate $P(n,k) = \frac{n!}{(n-k)!}$ permutations of $k$ elements, whereas join-cons patterns can be used to enumerate $C(n,k) = \frac{n!}{k!(n-k)!}$ combinations of $k$ elements.

```
matchAll [1,2,3] as list integer with _ ++ $x : _ ++ $y : _ -> (x,y) -- [[1,2],[1,3],[2,3]]
matchAll [1,2,3] as multiset integer with $x : $y : _ -> (x,y) -- [(1,2),(1,3),(2,1),(2,3),(3,1),(3,2)]
```

The descriptions of algorithms for which nested cons patterns for multisets are suitable become complicated in the traditional functional style. We can see that by just comparing the descriptions of the above two **matchAll** in functional programming.

However, pattern matching for multisets often appears in mathematical algorithms. Besides that, a much wider variety of patterns exist for multisets than lists. As a result, functions that correspond to patterns for multisets are not implemented as library functions because naming all these patterns is not practical. In functional programming so far, they are defined as a recursive function or combining several functions by users each time. It makes functional descriptions of mathematical algorithms complicated.

Thus, descriptions of these mathematical algorithms are the area where pattern-match-oriented programming demonstrates its full power. The rest of this paper discusses how we can describe this wide variety of patterns for multisets by just combining pattern constructs introduced in section 2.

### 3.3 Tuple Patterns with Sequential Not-Patterns for Comparing Collections

When describing algorithms, we often meet a situation to compare multiple data. A tuple pattern combined with not-patterns is especially useful for this purpose. For example, we can define difference by inserting a not-pattern into the definition of intersect in section 2.8.

```
difference xs ys = matchAll (xs, ys) as (multiset eq, multiset eq) with ($x : _, !(#x : _)) -> x
```

By changing the position of the not-pattern as !($x : _, #x : _), we can also describe a pattern that matches when no common element exists between the two collections.

We can write more complicated patterns by combining these patterns with a sequential pattern that allows us to apply a not-pattern to separate parts of the pattern simultaneously. For example, a pattern that matches when only one common element exists between the two collections is described below. A sequential pattern enables us to describe the pattern-matching process that first extracts one common element from the two collections, and after that checks that no common element exists between the remainder of the two collections. Sequential not-patterns often appear in mathematical algorithms, and we show an example again in section 4.1.

```
singleCommonElem = match (xs, ys) as (multiset eq, multiset eq) with
            [($x : @, #x : @),
             !($y : _, #y : _)] -> True
             _ -> False
```

We can combine a sequential pattern also with a loop pattern. For example, we can write a pattern that matches the common prefix of two lists with a sequential loop pattern.

```
match (xs, ys) as (list eq, list eq) with
 loop $i (1,$n)
  [($x_i : @, #x_i : @), [...]]
  !($y : _, #y : _)] -> map (\i -> x_i) [1..n]
```

### 3.4 Loop Patterns in Practice

Loop patterns are used for describing repetitions in a pattern. It is useful when we construct a complicated pattern by combining simple pattern constructors (section 3.4.1) and when the number of pattern variables that appear in a pattern changes by parameters (section 3.4.2). In such situations, very complicated recursion is necessary for describing algorithms. Loop patterns make the descriptions of these algorithms intuitive by confining recursion in a pattern. This section introduces such examples.

#### 3.4.1 Pattern Matching for Trees

This section demonstrates loop patterns by showing pattern matching for trees. The nodes of the trees in this section have an arbitrary number of children as XML, and they are handled as a multiset. A matcher for such a tree is defined as tree in section 2.8. We use this matcher in this section.

We describe patterns for a category tree of programming languages. treeData defines the category tree. For example, "Egison" belongs to the "pattern-match-oriented" category, and the "Dynamically typed" sub-category of the "Functional programming" category.

```
1  treeData =
2    Node "Programming language"
3      [Node "pattern-match-oriented" [Leaf "Egison"],
```

Functional Programming in Pattern-Match-Oriented Programming Style```
4     Node "Functional language"
5       [Node "Strictly typed" [Leaf "OCaml",Leaf "Haskell",Leaf "Curry",Leaf "Coq"],
6        Node "Dynamically typed" [Leaf "Egison",Leaf "Lisp",Leaf "Scheme",Leaf "Racket"]],
7     Node "Logic programming" [Leaf "Prolog",Leaf "Curry"],
8     Node "Object oriented" [Leaf "C++",Leaf "Java",Leaf "Ruby",Leaf "Python",Leaf "OCaml"]]
```

The **matchAll** expression below enumerates all categories to which a specified language belongs. A loop pattern is used to describe a pattern for this purpose because leaves can appear at an arbitrary depth. This pattern is interesting because the ellipsis pattern is not placed in the tail of the repeat pattern. The ability to choose the position of expansion of a repeat pattern allows us to apply the loop patterns to trees.

```
1 ancestors x t = matchAll t as tree string with
2   loop $i (1,$n)
3     (Node $c_i (... : _))
4     (Leaf #x) -> map (\i -> c_i) [1..n]
5
6 ancestors "Egison" treeData
7 -- [["Programming language","pattern-match-oriented"],"Programming language","Functional
   ↪ language","Dynamically typed"]]
```

It is also possible to enumerate all languages that belong to a specific sub-category. We can use a doubly-nested loop pattern for this purpose because it allows the sub-category to appear at an arbitrary depth. The following pattern matches all the languages that belong to a specified category. We used **matchAllDFS** for this enumeration to make the order of the languages in the result the same as the order with which they appear in the tree.

```
1 descendants x t = matchAllDFS t as tree string with
2   loop _ (1,_)
3     (Node _ (... : _))
4     (Node #x ((loop _ (1,_)
5              (Node _ (... : _))
6              (Leaf $y)) : _)) -> y
7
8 descendants "Functional language" treeData
9 -- ["OCaml","Haskell","Curry","Coq","Egison","Lisp","Scheme","Racket"]
```

Egison is more elegant than XML path language [33] for handling trees because we can describe the wide range of patterns by just combining a few simple pattern constructors and the loop patterns. In XML path, we would instead have to use the built-in ancestor command to enumerates all ancestors of a node, for example.

### 3.4.2 N-Queen Problem

This section introduces a more tricky example of nested loop patterns by introducing an $n$-queen solver in Egison. The $n$-queen problem is the problem of placing $n$ chess queens on an $n \times n$ board such that no queen can attack any of the other queens. In chess, a queen can attack other chess pieces on the same row, column, and diagonal.

Let us start by showing a program for solving the four queen problem. In this program, we represent the positions of the four queens with a list. The number of the $n$-th element represents the row number of the queen of the $n$-th line. The solution





must be a rearrangement of the list [1,2,3,4] because no two queens can be in the same line or row. Therefore, we pattern-match a collection [1,2,3,4] as a multiset of integers. The requirement that all two queens must not share the same diagonal is represented with conditions $a_1 \pm 1 \neq a_2$, $a_1 \pm 2 \neq a_3$, $a_2 \pm 1 \neq a_3$, $a_1 \pm 3 \neq a_4$, $a_2 \pm 2 \neq a_4$, and $a_3 \pm 1 \neq a_4$.

```
matchAll [1,2,3,4] as multiset integer with
  $a_1 :
  (and !#(a_1 - 1) !#(a_1 + 1) $a_2) :
   (and !#(a_1 - 2) !#(a_1 + 2) !#(a_2 - 1) !#(a_2 + 1) $a_3) :
    (and !#(a_1 - 3) !#(a_1 + 3) !#(a_2 - 2) !#(a_2 + 2) !#(a_3 - 1) !#(a_3 + 1) $a_4) :
     [] -> [a_1,a_2,a_3,a_4]
-- [[2,4,1,3],[3,1,4,2]]
```

We can use a doubly-nested loop pattern for generalizing this pattern for the $n$-queen solver. The index pattern variable i of the outer loop is referred to in the index range of the inner loop pattern for describing the difference of the repeat count of inner loop patterns. Also note that the values bound in the previously repeated pattern are referred as a_j in #(a_j - (i - j)) and #(a_j + (i - j)). Non-linearity of indexed pattern variables are effectively used for representing this pattern.

```
nQueens n =
  matchAll [1..n] as multiset integer with
    $a_1 :
    (loop $i (2,n)
       ((loop $j (1, i - 1)
           (and !#(a_j - (i - j)) !#(a_j + (i - j)) ...)
           $a_i) : ...)
       [] -> map (\i -> a_i) [1..n]

nQueens 4 -- [[2,4,1,3],[3,1,4,2]]
```

## 4 Pattern-Match-Oriented Programming in More Practical Situations

This section discusses how pattern-match-oriented programming changes the implementation of more practical algorithms and software.

### 4.1 SAT Solver

To see the effect of pattern-match-oriented programming for implementing practical algorithms, we implement a SAT solver. A SAT solver determines whether a given propositional logic formula has an assignment for which the formula evaluates to true. Input formulae for SAT solvers are often in *conjunctive normal form*. A formula in conjunctive normal form is a conjunction of clauses, which are disjunctions of *literals*. A literal is a formula whose form is $p$ or $\neg p$. For example, $(p \vee q) \wedge (\neg p \vee r) \wedge (\neg p \vee \neg r)$ is a formula in conjunctive normal form that has a solution; $p = \mathsf{False}$, $q = \mathsf{True}$, and $r = \mathsf{True}$.



**Functional Programming in Pattern-Match-Oriented Programming Style**

#### 4.1.1 The Davis-Putnam Algorithm

In our implementation, propositional logic formulae in conjunctive normal form are represented as a collection of collections of literals. We can pattern-match them as a multiset of multisets of literals because both ∧ and ∨ are commutative operators. Furthermore, we represent a literal as an integer. We represent positive and negative literals as positive and negative integers respectively: for example, $p$ and $\neg p$ are represented as 1 and $-1$, respectively. Therefore, the matcher for these formulae can be defined by simply composing matchers as multiset (multiset integer).

The program below shows the main part of our implementation of the Davis-Putnam algorithm[14]. The sat function takes a list of propositional variables and a logical formula, and returns **True** if there is a solution, otherwise returns **False**.

```
1  sat vars cnf = match (vars, cnf) as (multiset integer, multiset (multiset integer)) with
2    (_     , []) -> True
3    (_     , [] : _) -> False
4    (_     , ($x : []) : _) -> -- 1-literal rule
5      sat (delete (abs x) vars) (assignTrue x cnf)
6    ($v : $vs, !((#(negate v): _) : _)) -> -- pure literal rule (positive)
7      sat vs (assignTrue v cnf)
8    ($v : $vs, !((#v: _) : _)) -> -- pure literal rule (negative)
9      sat vs (assignTrue (negate v) cnf)
10   ($v : $vs, _) -> -- otherwise
11     sat vs (deleteClausesWith [v, negate v] cnf ++ resolveOn v cnf)
```

The first match clause states that the input formula has a solution when it is empty. The second match clause states that there is no solution when clauses include an empty clause. The third match clause represents *1-literal rule*. When the input formula includes a clause with a single literal, we can assign that literal **True** at once. The fourth match clause states that if there is a propositional variable that appears only positively, we can set the value of this literal **True** and remove the propositional variable of x from the variable list and the clauses that include this literal from the formula. For example, $(p \vee q) \wedge (\neg p \vee r) \wedge (\neg p \vee \neg r)$ contains a propositional variable $q$ only positively, so we can assign $q$ **True** and remove the first clause from the next recursion. The fifth match clause states the opposite of the fourth match clause. It removes the clauses that include this literal if there is a propositional variable that appears only negatively ($p$ in the above sample is such propositional variable). The final match clause applies the resolution principle. The resolveOn function collects all pairs of clauses $p \vee C$ and $\neg p \vee D$ (let $C$ and $D$ be a disjunction of literals), and returns new clauses $C \vee D$.

The above definition of sat describes all rules of the Davis-Putnam algorithm by fully utilizing pattern matching for multisets. In traditional functional languages, we need to call several library functions and define several helper functions to describe these conditional branches. We can compare this implementation with the same algorithm implemented in OCaml in [14].

#### 4.1.2 Pattern Matching for Resolution

We can elegantly define the resolveOn function using a sequential not-pattern.

First, let us show a naive implementation of resolveOn. The resolveOn function is defined with a single **matchAll** expression as follows.





```
resolveOn v cnf = matchAll cnf as multiset (multiset integer) with
  (#v : $xs) : (#(negate v) : $ys) : _ -> unique (filter (\c -> not (tautology c)) (xs ++ ys))
```

The pattern for enumerating the pair of clauses $p \vee C$ and $\neg p \vee D$ is described simply utilizing pattern-matching for a multiset of multisets. Note that the above resolveOn removes tautological clauses by using the tautology predicate in the body of the match clause.

We can remove this use of the tautology predicate by using a sequential not-pattern discussed in section 3.3. The sequential pattern is effectively used to describe that the literal x appeared in xs does not appear negatively in ys.

```
resolveOn v cnf = matchAll cnf as multiset (multiset integer) with
  [(#v : (and @ $xs)) : (#(negate v) : (and @ $ys)) : _,
   !($x : _, #(negate x) : _)] -> unique (xs ++ ys)
```

### 4.1.3 Separating Two Kinds of Loops

The SAT solver presented in this section is an important sample in the sense that it is the only sample that contains a loop that we cannot remove by pattern-match-oriented programming. This unremovable loop is the recursion of the sat function. This recursion is essential for narrowing the search tree. This narrowing is impossible by simple backtracking. On the other hand, all the other loops that can be implemented in backtracking algorithms are pushed into the patterns. In the traditional style, we usually describe the 1-literal rule and pure literal rules by combining several recursive functions such as find, partition, subtract and intersect [14]. Thus, pattern-match-oriented programming increases the readability of practical algorithms.

### 4.2 Computer Algebra

As an application of pattern-match-oriented programming, we have developed a computer algebra system [10]. We can implement a pattern-matching engine for mathematical expressions in a short program by defining a matcher for mathematical expressions.

```
1  mathExpr = matcher
2        Div $ $ as (mathExpr, mathExpr) with
3          Div $x $y -> [(x, y)]
4          $tgt -> [(tgt, 1)]
5        Poly $ as multiset mathExpr with
6          $tgt -> [tgt]
7        Term $ $ as (integer, multiset (mathExpr, integer)) with
8          Term $c $xs -> [(c, xs)]
9        App $ $ as (mathExpr, list mathExpr) with
10         App $f $args -> [(f, args)]
11       Sym $ as string with
12         Sym $name -> [name]
13       $ as something with
14         $tgt -> [tgt]
```



**Functional Programming in Pattern-Match-Oriented Programming Style**

The matcher for mathematical expressions is used for implementing programs for simplifying mathematical expressions. For example, a function that simplifies a mathematical expression by reducing $\cos^2(\theta) + \sin^2(\theta)$ to 1 can be implemented as follows.

```
simplifyCosAndSinInPoly poly =
  match poly as mathExpr with
    Poly (Term $n ((App (Sym #"cos") [$x]), 2):$y) : Term #n ((App (Sym #"sin") [#x], 2):#y) : r) ->
      simplifyCosAndSinInPoly (n * (prod (map power y)) + (sum r))
    _ -> poly
```

The definition of a matcher for mathematical expressions is simple compared with the pattern-matching engines of the other computer algebra systems. As a result, the implementation of the whole computer algebra system is also compact and straightforward; therefore, this computer algebra system is easily extensible. This extensibility allows us to experiment with new features easily.

This extensibility is a significant advantage in the future of computer algebra systems in the field of which there are still notations that are popular among researchers of mathematics but cannot be used in programs. There are many possibilities of research for extending computer algebra systems to support these notations. The extensibility of computer algebra systems will help us advance this research.

Such work has already been done by Egi, an author of this paper. He has developed a natural method for importing tensor index notation into programming [7]. Thanks to this work, Egison became an appropriate computer algebra system for describing formulae of differential geometry. We are now preparing a paper whose results are calculated using Egison collaborating with researchers of differential geometry.

### 4.3 Database Query Languages

Egison pattern matching can provide a unified query language for various kinds of databases. For example, let us consider a database of a social network service and a query to list all users who are followed by the user whose name is `"Egison_Lang"` but who do not follow this user. This query can be written with **matchAll** as follows using Egison pattern matching. This query pattern-matches the tables of a relational database as sets.

```
matchAll (users, follows, users) as (set user, set follow, set user) with
  ((and (Name #"Egison_Lang") (ID $uid)) : _,
   (and (FromID #uid) (ToID $fid)) : !((and (FromID #fid) (ToID #uid)) : _),
   (and (ID #fid) (Name $fname)) : _) -> (fid, fname)
```

The above **matchAll** expression matches a tuple of the user table (`users`), the follow table (`follows`), and the user table. Each table is pattern-matched as a set. The second line pattern-matches the user table. `Name` and `ID` are pattern constructors to access the column of a record. The pattern for the user table in the second line matches the user whose name is `"Egison_Lang"` and `$uid` is bound to the user ID of that user. The third line pattern-matches the follow table. `FromID` and `ToID` are pattern constructors to match the IDs of follower and followee. The user of `FromID` follows the user of `ToID`.





IDs of the users who do not follow back the user whose ID is uid is pattern-matched using a not-pattern. The fourth line pattern-matches the user table again to get the name of the user whose ID is fid and returns the tuple (fid, fname).

The conciseness of the queries is an important advantage of Egison over SQL [5]. For example, the same query described in SQL is more complicated. We need to write all conditions in WHERE clauses instead of a non-linear pattern and a sub-query instead of a not-pattern. A query in the pattern-match-oriented style can be interpreted by reading it once from left to right in order, whereas one in SQL cannot.

```
1  SELECT DISTINCT ON (user.name) user.name
2    FROM user AS user1, follow AS follow1, user AS user2
3    WHERE user1.name = 'Egison_Lang' AND follow1.from_id = user1.id AND user2.id = follow1.to_id
4    AND NOT EXISTS
5      (SELECT '' FROM follow AS follow2
6       WHERE follow2.from_id = follow1.to_id AND follow2.to_id = user1.id
```

List comprehensions also work as a sophisticated query language for relational databases [28]. The above query can be simply expressed also by list comprehensions. The advantage of Egison to list comprehensions as a query language is its generality. Egison can be used to express queries not only for relational databases but also for XML and graph databases. XML path language [33] and graph query languages [18, 21, 24] only focus on handling their target data structures and have many built-in functions to handle various patterns. On the other hand, Egison pattern-matching system allows users to describe various patterns for various data types in a unified manner with a small number of pattern constructors.

## 5 Related Work

This section reviews the history of pattern-match extensions (section 5.1) and list comprehensions (section 5.2) as another approach to remove explicit recursions from programs.

### 5.1 History of Pattern-Match Extensions

When pattern matching first appeared, pattern matching could only be applied to the specific types of algebraic data types [3]. Huge efforts have been conducted to remove this limitation [4, 15]. Pattern matching has evolved by bringing the programs that were written in the body of match clauses into the definitions of patterns. This importation has gradually proceeded. As a result, state-of-the-art work allows us pattern matching for non-free data types. This section reviews this evolution by seeing what happened in each decade.

#### 5.1.1 1980s: Spread of Pattern Matching and Invention of Views

From this decade, functional languages with user-defined algebraic data types and pattern matching for them became common. Miranda by Turner [30] and Haskell [15] were the most popular among these languages, and the first pattern-match extensions





for widening the target of pattern matching beyond algebraic data types were designed on them.

Miranda's laws [25, 26] and Wadler's views [32] are earlier such research. They discarded the assumption that one-to-one correspondence should exist between patterns and data constructors. They enable pattern matching for data types whose data have multiple representation forms. For example, Wadler's paper on views [32] present pattern matching for complex numbers that have two different representation forms: Cartesian and polar. However, their expressiveness is not enough for representing patterns for non-free data types. They support neither non-linear patterns nor pattern matching with multiple results. Views are supported as a GHC extension in Haskell [31]. Views are implemented also in Racket [27].

At the same time, more expressive pattern matching is explored by Queinnec [20], who proposed an expressive pattern matching for lists. Though this proposal is specialized to lists and not extensible, the proposed language supports the cons and the join patterns, non-linear pattern matching with backtracking, matchAll, not-patterns, and recursive patterns. His proposal achieves almost perfect expressiveness for patterns of lists and allows the pattern-match-oriented definition of the basic list processing functions. For example, the following member definition is presented in Queinnec's paper [20].

```
member ?x (??- ?x ??-) -> true
member ?x ?-          -> false
```

### 5.1.2 1990s and 2000s: Exploration for Expressive Patterns

Following the pattern-match extensions in the previous decade, several new pattern-match extensions for extending the target range of pattern matching have been proposed by several researchers. We review these proposals in this section.

Erwig's active patterns [11] are an attempt to extend the expressiveness of patterns beyond Wadler's views. Active patterns also allow users to customize the pattern-matching algorithm for each pattern. Add' in the following program is a pattern constructor of active patterns. Add' extracts an element that is identical with the first argument of Add' from the target collection.

```
pat Add' (x,_) = Add (y,s) => if x == y then Add (y,s) else let Add' (x,t) = s in Add (x, Add (y, t)) end
```

Using the above Add', we can define the member function as follows, hiding the recursion as the pattern-match-oriented definition of map we presented in section 1.

```
fun member x (Add' (x,s)) = true
  | member x s = false
```

However, the expressiveness of active patterns is still limited. Active patterns do not support pattern matching with multiple results: Add' can take only a value and cannot take a pattern variable as its first argument. Non-linear patterns exhibit their full ability when they are combined with pattern matching with backtracking.

Tullsen's first class patterns [29] are another extension of views. First class patterns support pattern matching with multiple results. In first class patterns, we can define pattern constructors that have multiple decompositions. However, the expressiveness





of first class patterns is still limited because it does not support non-linear patterns. Non-linear pattern matching is a necessary feature for describing useful patterns for non-free data types.

Functional logic programming is an independent approach for pattern matching against non-free data types. Curry [13] by Hanus is the most popular functional logic programming language. Curry supports both properties: non-linear patterns and pattern matching with multiple results. Therefore, we can write expressive patterns for non-free data types in Curry.

However, Curry does not intend to utilize non-linear pattern matching with multiple results fully. For example, non-linear pattern matching is not efficiently executed in Curry. Theoretical time complexities depend on the size of patterns.[5] The reason for the difference in time complexities between these patterns is that Curry transforms non-linear patterns into pattern guards [1, 2, 12]. Pattern guards are applied after enumerating all pattern-matching results. As a result, considerable unnecessary enumerations often occur before the application of pattern guards.

Moreover, Curry does not provide a special syntactic construct for handling multiple pattern-matching results. Curry provides findall for handling multiple unification results. However, if we use findall for pattern matching, the program gets more complicated than the functional approach. For example, Curry program that defines the map function in the pattern-match-oriented style is as follows.

```
map f xs = findall (\y -> let x free in (_ ++ (x : _)) =:= xs & f x =:= y)
```

### 5.1.3 2010s: Toward a Unified Theory of Pattern-Match-Oriented Programming

In this decade, a unified theory for practical pattern matching for non-free data types has been pursued. Egison [9] proposed by the same authors of this paper is such research. The research listed and organized the properties for practical pattern matching for non-free data types. They proposed three criteria in their paper. The criteria are as follows: (1) Efficiency of the backtracking algorithm for non-linear patterns; (2) Ad-hoc polymorphism of patterns; (3) Extensibility of pattern matching.

## 5.2 List Comprehensions

List comprehensions [19] are another approach to hide explicit recursions. For example, list comprehensions also allow us to define map without explicit recursion:

```
map f xs = [ f x | x <- xs ]
```

However, list comprehensions are too specialized to enumerate elements of a list and do not allow us to describe complex enumerations as concisely as pattern-match-oriented programming. We can summarize the advantages of pattern-match-oriented programming against list comprehensions as follows:

1. Pattern-match-oriented programming requires less local variables;

---

[5] Our previous paper [9] shows benchmark results.





2. Pattern-match-oriented programming is more expressive thanks to ad-hoc polymorphism of patterns and Egison specific pattern constructs such as loop patterns and sequential pattern;
3. Pattern matching can be used to describe conditional branches.

The second and third advantages are obvious. Therefore, in the rest of this section, we focus on the first advantage.

For illustrating the first advantage, we write a program that enumerates all the two combinations of elements in list comprehensions. We described the same program in pattern-match-oriented programming style in section 3.2. tails is a function that returns all the suffixes of the argument list.

```
comb2 xs = [ (x,y) | x:ys <- tails xs, y <- ys ]
```

ys is a variable that is necessary in list comprehensions though it is unnecessary in pattern-match-oriented programming. Such variables that are necessary only in list comprehensions appear when the pattern is nested. As a result, nested join-cons patterns for lists (e.g. unique in section 3.1.2) and nested cons patterns for multisets (e.g. an N-queen solver in section 3.4.2 and a SAT solver in section 4.1) cannot be described in list comprehensions as concisely as pattern-match-oriented programming.

# 6 Conclusion

This paper has presented programming techniques that replace explicit recursions for traversing search trees with intuitive patterns and allow programmers to concentrate on the description of the essential parts of algorithms that reduce the computational complexities of the algorithms. This paper listed many algorithms that can be represented more elegantly in this way. These include many very basic list processing functions to some larger more practical algorithms.

We believe the development of these programming techniques that enable the more intuitive representation of algorithms extends the limit on the complexity of software that we can practically implement and has the potential to accelerate research of computer science as a whole. We focused on the data abstraction for non-free data types in this paper. However, there must be more data types waiting to be abstracted. We hope this paper leads to the further evolution of pattern matching and the future progress of data abstraction.

**Acknowledgements** We thank Kentaro Honda, who is the first real user of Egison, for his contribution to some sample programs in this paper. We thank Pierre Imai, Matthew Roberts, Paul Harvey, and Masami Hagiya for helpful feedback on the earlier versions of the paper. Finally, we greatly thank the anonymous reviewers for their helpful reviews.

## A  Creating Your Own Matchers

There are data types whose matchers cannot be defined by just composing the existing matchers. For example, matchers for lists and multisets are such matchers. This section explains how to define matchers from scratch through examples. We omit an explanation of the formal semantics of matchers and the pattern-matching algorithm inside Egison and instead focus on an explanation of programming techniques used for defining matchers. For the detailed description of the formal semantics, please refer to the original paper of Egison pattern matching [9].

Appendix A.1 explain how to create a matcher by showing a definition of the multiset matcher. We show that the pattern-match-oriented programming techniques presented in this paper are useful also for matcher definitions. Appendix A.2 introduces a programming technique specific to matcher definitions by showing a definition of the sorted-list matcher.

### A.1  Multiset Matcher

Matcher definition is the most technical part in pattern-match-oriented programming. This section explains the method for defining a matcher by showing a matcher definition for multisets. The program below shows a matcher definition for multisets. The multiset matcher is the most basic nontrivial matcher. This section explores the detail of this definition.

```
multiset a = matcher
       [] as () with
         \tgt -> case tgt of
              [] -> [()]
              _ -> []
       $ : $ as (a, multiset a) with
         \tgt -> matchAll tgt as list a with
              $hs ++ $x : $ts -> (x, hs ++ ts)
       #$val as () with
         \tgt -> match (val, tgt) as (list a, multiset a) with
              ([], []) -> [()]
              ($x : $xs, #x : #xs) -> [()]
              (_, _) -> []
       $ as something with
         \tgt -> [tgt]
```

A matcher is defined using the matcher expression. The matcher expression is a built-in syntax of Egison. matcher takes a collection of *matcher clauses*. A matcher clause is a triple of a *primitive-pattern pattern*, a *next-matcher expression*, and a *next-target expression*.





A matcher is a kind of function that takes a pattern and target, and returns lists of the next *matching atoms*. A matching atom is a triple of a pattern, target, and matcher. A primitive-pattern pattern matches a pattern. Patterns that match with *pattern holes* ($ inside primitive-pattern patterns) are next patterns. A next-matcher expression returns next matchers. A next-target expression is a function that takes a target and returns a list of next targets. A matcher generates a list of the next matching atoms by combining next patterns, next matchers, and a list of next targets.[6]

The multiset matcher has four matcher clauses. The first matcher clause handles a nil pattern, and it checks whether the target is an empty collection or not. The second matcher clause handles a cons pattern. The third matcher clause handles a value pattern. This matcher clause defines the equality of multisets. The fourth matcher clause handles the other patterns for multiset: a pattern variable and wildcard.

First, we focus on the second matcher clause. The primitive-pattern pattern of the second matcher clause is $:$, and the next matcher expression is (a, multiset a). It means two arguments of the cons pattern are next patterns and they are pattern-matched using the a and multiset a matchers, respectively. a is an argument of multiset and the matcher for inner elements of a multiset. In the next target expression, a simple join-cons pattern is used to decompose a target collection into an element and the rest collection. For example, when the target is a collection [1,2,3], this next-target expression returns [(1,[2,3]),(2,[1,3]),(3,[1,2])]. Each tuple of the next targets is pattern-matched using the next patterns and the next matchers recursively. For example, 1 and [2,3] are pattern-matched using the a and multiset a matcher with the first and the second argument of the cons pattern, respectively.

Next, we focus on the third matcher clause. This matcher clause is as technical as the second matcher clause. The primitive-pattern pattern of this matcher clause is #$val. It is called a *value-pattern pattern*. A value-pattern pattern matches a value pattern. This matcher clause compares the content of a value pattern (val) and a target (tgt) as multisets. The **match** expression is used for this comparison. Interestingly, tgt is recursively pattern-matched as multiset a.

The first and the third match clauses of this **match** expression are simple. The first match clause states that it returns [()] when both val and tgt are empty. This return value means pattern matching for the value pattern succeeded. The third match clause states that it returns [] if pattern matching for the patterns of both the first and the second match clause failed. This return value means pattern matching for the value pattern failed.

The second match clause is the most technical part of this **match** expression. The value pattern #xs is recursively pattern-matched using this matcher clause itself. The

---

[6] In Egison, pattern matching is implemented as reductions of stacks of matching atoms. Each list of the next matching atoms returned by a matcher is pushed to the stack of matching atoms. As a result, a single stack of matching atom is reduced to multiple stacks of matching atoms in a single reduction step. Pattern matching is recursively executed for each stack of matching atoms. When a stack becomes empty, it means pattern matching for this stack succeeded.





collection xs is one element shorter than tgt. Therefore, this recursion finally reaches the first or the third match clause if val and tgt are finite.

Finally, let us also explain the fourth matcher clause. This matcher clause creates the next matching atom by just changing the matcher from multiset a to something.

### A.2 Matcher for Sorted Lists

Modularization of pattern-matching algorithms by matchers not by patterns enables polymorphic patterns. However, its merit extends beyond polymorphic patterns; matchers enable descriptions of more efficient pattern matching keeping patterns concise. The reason is that pattern matching against patterns inside matcher definitions allows us to describe more detailed pattern-matching algorithms. This section shows such an example, a matcher for sorted lists.

The program that used a doubly-nested join-cons pattern for enumerating pairs of prime numbers whose forms are $(p, p + 6)$ gets slower when the number of the enumerating prime pairs gets larger. The reason is that the program enumerates all the combinations of prime numbers. For example, the program tries to match all the pairs such as $(3, 5), (3, 7), (3, 11), (3, 13), (3, 17), (3, 19)$, and so on. However, we should avoid enumerating the pairs after $(3, 11)$, the first pair whose difference is more than 6. This is because it is obvious that the difference between all the pairs after $(3, 11)$ are more than 6.

```
take 10 (matchAll primes as sortedList integer with
     _ ++ $p : (_ ++ #(p + 6) : _) -> (p, p + 6))
-- [(5,11),(7,13),(11,17),(13,19),(17,23),(23,29),(31,37),(37,43),(41,47),(47,53)]
```

We can avoid this unnecessary search by creating a new matcher that is specialized for sorted lists. We can define such a matcher by adding a matcher clause with the primitive-pattern pattern $ ++ #$px : $ to the list matcher as shown below. This matcher clause improves the theoretical time complexity of the above pattern from $O(n^2)$ to $O(n)$.

```
1 sortedList a =
2   matcher
3    $ ++ #$px : $ as (sortedList a, sortedList a)
4      \tgt -> matchAll tgt as list a with
5           loop $i (1, $n) ((and ?(\x -> x < px) $h_i) : ...) (#px : $ts) ->
6             (map (\i -> h_i) [1..n], ts)
7    ...
```

Note that this method is only applicable to Egison that modularizes pattern-matching methods for each matcher, not for each pattern. The reason is that we need to match patterns whose form is _ ++ #x : _ as mentioned above. If pattern-matching methods are modularized for each pattern, we need to introduce a new pattern constructor joinCons ... ... that is equivalent to _ ++ ... : ... for this purpose.





# B Poker Hand Pattern Matching

Pattern matching for poker hand is a simple example of the application of Egison pattern matching. All poker hands are described in a single pattern utilizing non-linear pattern matching for a multiset. When the creator of Egison, an author of this paper, designed the syntax of Egison, he designed the syntax of patterns to make this poker hand program as concise as possible.

```
1  poker cs =
2    match cs as multiset card with
3      (card $s $n):(card #s #(n - 1)):(card #s #(n - 2)):(card #s #(n - 3)):(card #s #(n - 4)):[] ->
4        "Straight flush"
5      (card _ $n):(card _ #n):(card _ #n):(card _ #n):_ ->
6        "Four of kind"
7      (card _ $m):(card _ #m):(card _ #m):(card _ $n):(card _ #n):[] ->
8        "Full house"
9      (card $s _):(card #s _):(card #s _):(card #s _):(card #s _):[] ->
10       "Flush"
11     (card _ $n):(card _ #(n - 1)):(card _ #(n - 2)):(card _ #(n - 3)):(card _ #(n - 4)):[] ->
12       "Straight"
13     (card _ $n):(card _ #n):(card _ #n):_ ->
14       "Three of kind"
15     (card _ $m):(card _ #m):(card _ $n):(card _ #n):_ ->
16       "Two pair"
17     (card _ $n):(card _ #n):_ ->
18       "One pair"
19     _ -> "Nothing"
```

The card matcher is defined as an algebraic data matcher as follows:

```
algebraicDataMatcher card = Card suit integer
algebraicDataMatcher suit = Diamond | Clover | Heart | Spade
```

# C Graph Pattern Matching

This section demonstrates pattern matching for graphs as sets of edges and adjacency graphs, respectively.

## C.1 Graphs as Sets of Edges

In this section, we pattern-match a graph as a set of edges. We can define a matcher and graph data as follows.

```
1  graph = set edge
2  algebraicDataMatcher edge = Edge integer integer
3
4  graphData = [Edge 1 2,Edge 2 1,Edge 2 3,Edge 2 4,Edge 3 4,Edge 4 5,Edge 4 6,Edge 4 7,
5               Edge 5 4,Edge 5 6,Edge 5 7,Edge 6 4,Edge 6 5,Edge 6 7,Edge 7 4,Edge 7 5,Edge 7 6,
6               Edge 7 8,Edge 9 10,Edge 10 7]
```





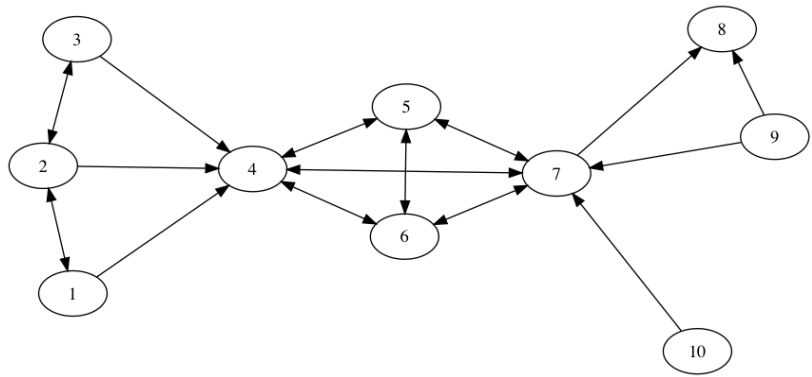

**Figure 1** Visualization of graphData.

This sample graph is visualized in figure 1. This section lists several pattern-matching examples against this graph. Various patterns for graphs can be described using techniques introduced in this paper.

A pattern for listing all nodes that are accessible from s in two edges is described as follows.

```
1  let s = 1 in
2    matchAll graphData1 as graph with
3      Edge (and #s $x_1) $x_2 : Edge #x_2 $x_3 : _ -> x
4  -- [1,3,4,5,6,7]
```

A pattern for listing all nodes that posses an edge from s but not to s is described utilizing a not-pattern effectively.

```
1  let s = 1 in
2    matchAll graphData1 as graph with
3      Edge #s $x : !(<Edge #x #s> : _) -> x
4  -- [4]
```

A pattern for listing all routes from s to e is defined with a loop pattern. Egison allows users to use the let expression inside a pattern. The let expression is used to bind s to $x_1. Thanks to this let, we can describe elegantly an initial condition for $x_1 for the loop pattern.

```
1  let (s, e) = (1, 8) in
2    matchAll graphData as graph with
3      let x_1 = s in
4        loop $i (2,$n)
5          ((Edge #x_(i - 1) $x_i) : ...)
6          ((Edge #x_(n - 1) (and #e $x_n)) : _) -> map (\i -> x_i) [1..n]
7  -- [[1,4,7,8], ...]
```

A pattern for finding all cliques whose size is $n$ is given in the following way. A double-nested loop pattern can be used for that purpose.

```
1  let n = 4 in
2    matchAll graphData2 as graph with
3      Edge $x_1 $x_2 : loop $i (3,n)
4             (Edge #x_1 $x_i : loop $j (2, i - 1)
```





```
5                         (Edge #x_j #x_i) : ...)
6                            ...)
7              _ -> map (\i -> x_i) [1..n]
8  -- [[4,5,6,7],...]
```

In this section, we demonstrated pattern matching only for a directed graph. However, we can also define a matcher for undirected graphs using a matcher for edges that identifies `Edge a b` and `Edge b a`. The matcher for undirected edges are defined as a user-defined matcher, which we explain in appendix A. [7]

### C.2 Adjacency Graphs

This section demonstrates pattern matching for a weighted adjacency list. As shown in the program below, a matcher for a weighted adjacency list can be simply defined by composing matchers. `graphData` in the program below represents an airline network by a weighted adjacency list. The integers in `graphData` are the costs of time by hours to move between two cities.

```
1  graph = multiset (string,multiset (string,integer))
2
3  graphData =
4    [("Berlin", [("New York", 14), ("London", 2), ("Tokyo", 14), ("Vancouver", 13)]),
5     ("New York", [("Berlin", 14), ("London", 12), ("Tokyo", 18), ("Vancouver", 6)]),
6     ("London", [("Berlin", 2), ("New York", 12), ("Tokyo", 15), ("Vancouver", 10)]),
7     ("Tokyo", [("Berlin", 14), ("New York", 18), ("London", 15), ("Vancouver", 12)]),
8     ("Vancouver", [("Berlin", 13), ("New York", 6), ("London", 10), ("Tokyo", 12)])]
9
10 -- List all routes that visit all cities exactly once and return to Berlin.
11 trips =
12   let n = length graphData in
13     matchAll graphData as graph with
14       (#"Berlin" (($s_1,$p_1) : _)) : loop $i (2, n - 1)
15                         ((#s_(i - 1), ($s_i,$p_i) : _) : ...)
16                         ((#s_(n - 1), (and #"Berlin" $s_n, $p_n) : _) : []) ->
17       (sum (map (\i -> p_i) [1..n]), map (\i -> s_i) [1..n])
18
19 head (sortBy (\(_, x), (_, y) -> compare x y)) trips)
20 -- (["London","New York","Vancouver","Tokyo","Berlin"], 46)
```

The above **matchAll** expression lists all routes from Berlin that visit all the cities exactly once and return to Berlin. This pattern can be used to solve the traveling salesman problem. A non-linear loop pattern is used to represent the pattern.

There are several graph database query languages [18, 21, 24]. The advantage of Egison over these query languages is its generality. Egison does not focus on pattern matching for graphs. Instead, Egison allows users to describe various patterns by just combining non-linear loop patterns and a small number of simple pattern constructors.

---

[7] The matcher for unordered pairs defined in [9] is such a matcher.



**Functional Programming in Pattern-Match-Oriented Programming Style**

## About the authors

**Satoshi Egi** is a research scientist at Rakuten Institute of Technology, a research organization of Rakuten, Inc. He is also a PhD student of the University of Tokyo. His interest is in intuitive representation of mathematics and automated reasoning. He often implement his ideas in the Egison programming language. He received IPSJ Software Japan Award in 2015, Japan OSS Encouragement Award in 2015, and Mitoh Super Creator Award in 2012. satoshi.egi@rakuten.com

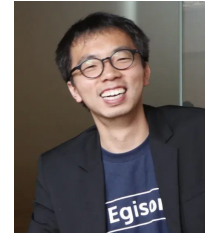

**Yuichi Nishiwaki** Yuichi Nishiwaki is a PhD student in Hagiya laboratory at the University of Tokyo. He has been working on developing theories and applications of programming language, particularly from the mathematical perspective. His research often involves methods from semantics, logic, and category theory. nyuichi@is.s.u-tokyo.ac.jp

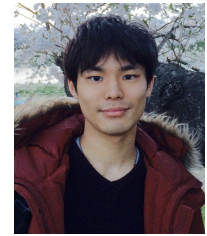